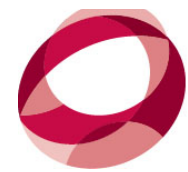

# Intelligent Infrastructure for Smart Agriculture: An Integrated Food, Energy and Water System

Shashi Shekhar
University of Minnesota

Joe Colletti
Iowa State University

Francisco Muñoz-Arriola
University of Nebraska-Lincoln

Lakshmish Ramaswamy
The University of Georgia

Chandra Krintz
University of California, Santa Barbara

Lav Varshney
University of Illinois at Urbana-Champaign

Debra Richardson
University of California, Irvine

**Abstract:** Agriculture provides economic opportunity through innovation; helps rural America to thrive; promotes agricultural production that better nourishes Americans; and aims to preserve natural resources through healthy private working lands, conservation, improved watersheds, and restored forests. From agricultural production to food supply, agriculture supports rural and urban economies across the U.S. It accounts for 10% of U.S. jobs and is currently creating new jobs in the growing field of data-driven farming. However, U.S. global competitiveness associated with food and nutrition security is at risk because of accelerated investments by many other countries in agriculture, food, energy, and resource management. To ensure U.S. global competitiveness and long-term food security, it is imperative that we build sustainable physical and cyber infrastructures to enable self-managing and sustainable farming. Such infrastructures should enable next generation precision-farms by harnessing modern and emerging technologies such as small satellites, broadband Internet, tele-operation, augmented reality, advanced data analytics, sensors, and robotics.

## 1. Agriculture: A Vital Economic Engine and National Resource

Agriculture provides economic opportunity through innovation, helps rural America to thrive, promotes agricultural production that better nourishes Americans, as well as provides new jobs and aims to preserve natural resources through healthy private working lands, conservation, improved watersheds, and restored forests. This vital sector affects each American and is an economic engine that provides approximately 1 in 10 U.S. jobs. U.S. agricultural productivity and profitability are the envy of the world. Notably, many new jobs are appearing in the fast-growing area of data-driven farming[1]. However, U.S. global competitiveness associated with food and nutrition security is at risk because of accelerated investments in agriculture, food, energy, and resource management by countries such as China, Brazil, and India.

Technological innovation in agriculture, food manufacturing, energy production, and water conservation by the private and public sectors have been key drivers of our international competitiveness. A renewed private-public effort is needed to develop and deploy advanced technologies that will ensure food and nutrition security; address workforce, malnutrition and obesity; foster energy independence; manage critical natural resources, especially water; and improve our ability to adapt to environmental and market shocks that jeopardize food, energy, and water security.

## 2. The Potential for Intelligent Agriculture Infrastructure

An intelligent agriculture infrastructure that leverages private development and public R&D is the key to addressing these grand challenges and increasing our competitive position globally. [2] Interconnecting existing and new models (e.g., crop, soil, water and weather), creating public-private spaces for data sharing, and developing technologies distributed across all scales of food, energy, and water systems will create data-based assets that can be used to optimize agricultural productivity and the food pipeline all the way to consumer behavior and waste management, while increasing jobs, wages, and wealth creation opportunities in both rural and urban America. Data-based assets detailed in Box 1 include raw data, processed data, tools for real-time decision-making, and tools for models that stand alone and hopefully in the future will be interconnected. Further, these data-based assets need to be generated and shared across the private and public sectors.

---

[1] The Seeds of Innovation – Big Data Reshaping U.S. Agriculture, US Chamber of Commerce, March 2014, https://www.uschamberfoundation.org/blog/post/seeds-innovation-big-data-reshaping-us-agriculture/34140

[2] Mynatt et al. (2017) "A National Research Agenda for Intelligent Infrastructure" CCC Led Whitepapers http://cra.org/ccc/resources/ccc-led-whitepapers/, last accessed April 12, 2017.

> **Box 1. Data-based Assets**
> - Raw and processed data from farm equipment, field sensors, UAS, weather stations, and satellite resources. Some data originate with the farmers and agribusiness and other data originate in the public sector. Data typically captured include production inputs, fertility and moisture content of soil, topographic attributes, temperature, relative humidity, wind, precipitation, solar input, images of crops at various growth stages, crop yield, and crop health/quality traits (e.g., chlorophyll, moisture content).
> - Decision support tools that use raw and/or processed data to assist the farmers or agribusiness in real time with respect to input and crop management needs (seed, fertilizers, pesticides, soil amendments [e.g., lime], cultivation, and harvest).
> - Decision support tools that use raw and/or processed data in prediction models to assist the farmers with spatially-connected yield estimates and forecasting cost of production, profitability, and return on investment.

Over last few decades, computing research has unleashed game-changing capabilities in precision agriculture, including enabling farmers to optimize farm returns, reduce unnecessary applications of fertilizers and pesticides, preserve natural resources, and contend with impending weather events. Precision agriculture represents a holistic view of agriculture as an integrated food, energy and water system. Farmers monitor crop or animal growth and productivity, while sensing the efficient use of water and energy resources. Precision agriculture uses a blend of computing components such as global positioning systems, sensors to monitor soil and crop health, computerized map visualization to understand inter- and intra-field variability, spatial and temporal databases to collect and query farm data, spatial statistical analysis to delineate management zones, and spatial decision support systems to optimize yield while preserving natural and farm resources. These components and capabilities enable service-based operations and decision making at multiple levels, namely, descriptive, prescriptive, predictive and proactive levels:

- <u>Descriptive</u>: For precision agriculture and high throughput phenotyping applications, data collection aims to characterize spatial and temporal variability in soil, crop and weather characteristics and identify stressors and traits that need better management.
- <u>Prescriptive</u>: Using collected data and associated maps of individual characteristics or traits, a prescriptive analysis is conducted to determine necessary farm management interventions.
- <u>Predictive</u>: Similarly, a predictive analysis that uses historic data as well as integrated soil, crop and weather models may forecast crop yield at the end of the season.
- <u>Proactive</u>: Proactive involves observations of crop development and stress on multiple farms over large regions and time scales. Data from these observations are pooled and mined to obtain relationships between site characteristics, weather and crop performance under a range of management conditions. These relationships can be used to customize management practices and seed selection to local conditions.

Ultimately, farmers want to maximize production/revenues while minimizing costs and use of resources. Data scientists and software engineers can collaborate with farmers, educators and researchers to create tools that optimize the use of resources. The creation of such tools evidences the need for continuous, solid and reliable production of data; the design of analytics that translate such data into information; and the ability to synthesize and deliver the right information in time and space.

## 3. Smart Agriculture Challenges
### 3.1. Smart Agriculture as an Integrated Food-Energy-Water System

Figure 1 shows five major activities – namely, food production, processing, distribution, consumption and waste management – for food system outcomes such as food security. Food production can be characterized by three broad categories of food: fish, meat, and crops (e.g., grains, vegetables, fruits).

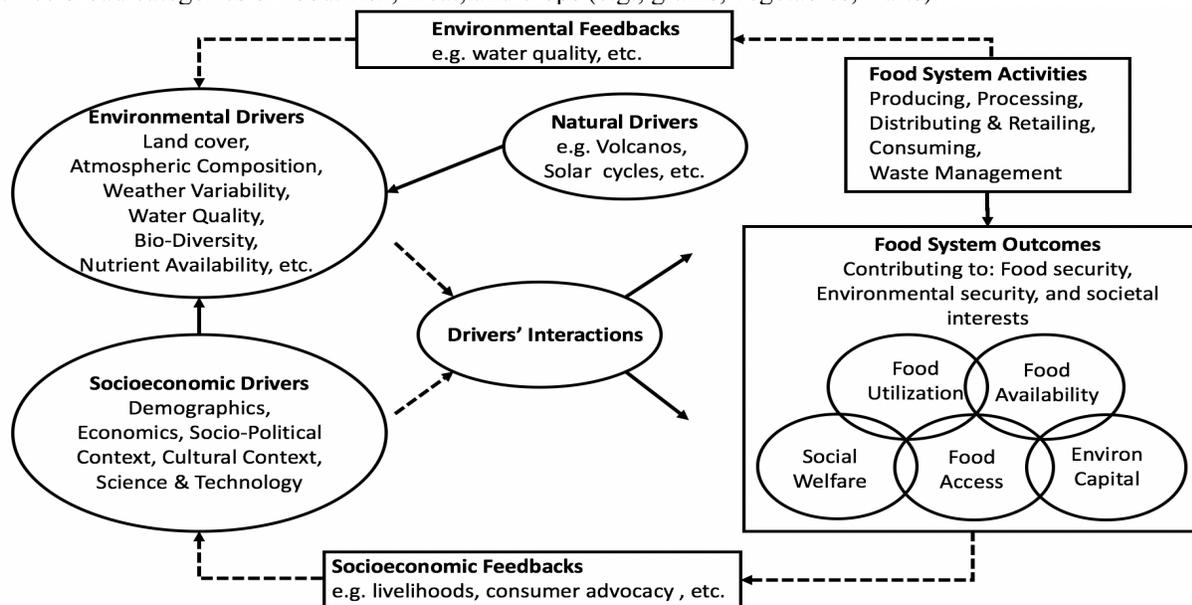

*Figure 1. Food systems and their drivers*[3]

It is not sufficient to consider the food system in isolation since many economic, social, and environmental drivers affect food security. In addition, the interactions among these drivers, activities and outcomes are complex. These drivers include weather/climate, land use change, urbanization, and population growth. As elements of such drivers, water and energy availability regulate sustainable development and infrastructure's resilience. Further, in a dynamic, yet complex system, effects of changing weather and climate, reflected in water scarcity and increasing energy demands, can influence food availability from its production to its delivery, and likely observe cumulative or magnified effects on communities with restricted economic power and access to food, water and energy resources[4].

In summary, agriculture is an integrated Food-Energy-Water system with many components and processes that regulate flows of mass, energy, and water across its components. Such flows are seen in water relocation to fulfill intra-seasonal and inter-annual deficits in agricultural working lands. At the same time, water relocation requires energy use and the adequate and timely availability of energy may represent the success of production or transport of food. Adverse economic factors such as an excessive food stock may deplete the price of produce or a climate spell may increase it in places with average climate conditions. Thus assessments of socioeconomic and physical vulnerability are essential to build resilient and sustainable integrated Food-Energy-Water systems.

### 3.2  Current Agriculture Infrastructure

Existing agriculture infrastructure may be classified by activities – namely, production, processing, distribution, consumption and waste management – with the exception of cross-cutting cyberinfrastructure and workforce development. For instance, food production infrastructure includes farm equipment such as tractors, combine-harvesters, irrigation infrastructure and duster airplanes. Support services for food production including germplasm[5] (e.g., seed banks), fertilizer production, and farm equipment manufacturing. In food processing, infrastructure includes food factories and processing plants. Food distribution infrastructure includes transportation systems (e.g., railroads, trucks, barges, ships, etc), supply chain and retail stores.

---

[3] Conceptualizing food sys. for global env. change research, P. Ericksen, Gl. Env. Change, 18:234-245, Elsevier, '08.
[4] Climate Change and Food Systems, S. Vermerulen, et al., Annual review of Env. & Resources, 37:195-222, '12.
[5] GRIN NPGS - Germplasm Resources Information Network, *http://www.ars-grin.gov*

In addition, agriculture also needs water and energy infrastructure. Water infrastructure includes pumping systems, built and natural water storage; monitoring networks, water governance and management. Energy infrastructure includes the electric grid, biofuel production, and fossil and carbon energy production.

Cyberinfrastructure is an important and cross-cutting component of existing agricultural infrastructure. It facilitates all phases of the data lifecycle, namely: (a) Data collection and citizen engagement via manual scouting, proximal and remote sensors and public and private investment in field monitoring networks, in-air sensor on UAVs, and in-space sensors on satellites; (b) Collection, curation, management, and long-term preservation of data for future discovery; (c) Model calibration/validation, data analytics, and synthesis involving interdisciplinary research between agronomists, hydrologists, physicists and computer and social scientists to generate and deliver better information; (d) Visualization and communication of information including the design of architectures as well as forms to better communicate information (e.g. social media). Examples of such cyberinfrastructure systems include the USDA VegScape[6], and the international Group on Earth Observations Global Agricultural Monitoring Initiative (GEOGLAM)[7] which use remotely sensed satellite imagery for monitoring major crops biweekly for end-of-season yield forecasts to enable timely interventions and reduce disruptions in global food supply. In addition, agriculture infrastructure also includes supply-chain management with all the organizations, people, activities, information, and resources involved in moving the products from the suppliers to the consumers.

### 3.3. Limitations of Current Agriculture Infrastructure

Current infrastructure has several limitations and challenges that need to be addressed to realize the vision of intelligent agriculture infrastructure. Social challenges include the aging workforce, labor shortage, and lack of engagement of urban communities. Currently, production of many crops is highly labor-intensive and the average age of the US farmer is about 60 years. The farming sector is projected to face severe shortages of skilled labor in the next decade. Environmental challenges include declining pollinator population (e.g., bees) in many areas.

There are also many technical challenges that need to be addressed. For example, current cyberinfrastructure is inadequate since the Internet bandwidth is severely limited in many farming areas and is easily overwhelmed by the data deluge from precision agriculture. The Global Positioning System (GPS) infrastructure used for positioning precision farming equipment are aging and increasingly vulnerable to jamming and spoofing. In addition, data sharing infrastructure is inadequate due to a lack of mechanisms to share agricultural data in a privacy-protected manner. The best-of-the-breed privacy protection technology, namely, differential privacy, is inadequate for agricultural data due to its spatiotemporal nature. Consequently, many farmers are reluctant to share data to allow for computing research due to legal, insurance, and privacy or other concerns. Furthermore, existing sensing infrastructure only allows infrequent (e.g., bi-weekly) and coarse resolution (e.g., 30 meter-pixels) monitoring via remote sensing satellites which introduces major delays in detection of adverse conditions and corrective actions. Furthermore, the level of automation in the farming sector varies significantly. For example, corn, wheat and pecan farms have higher levels of automation than fruit and vegetable farming and manual interventions are needed since these crops do not ripen at the same time and require periodical pickup. In addition, land management and fertilizer application decisions are made by farmers using manual procedures. Finally, there are also organizational challenges such as the lack of end-to-end visibility and transparency in the current supply-chain.

### 4. The Case for Federal Action for Smart Agriculture

Addressing these critical societal needs requires investment in intelligent infrastructure and computing research that concurrently increases economic competitiveness, intensifies food production, reduces resource use (e.g., land, water, and manual labor), and ensures long-term environmental viability and food safety. Table 1 (next page) lists many examples of infrastructure investment needs and opportunities. For example, tele-operation[8] may address the structural mismatch between farming areas facing labor shortage and other areas (e.g., old mining and manufacturing towns) with worker surplus. The tele-operation will also require investments in broadband Internet infrastructure for rural areas to support quick interaction between remote workers and farm equipment. Furthermore, investment in augmented (or virtual) reality infrastructure (e.g., farm-based video games similar to SimCity and flight simulator) may help engage the next generation even from urban areas in farm careers. Another major

---

[6] VegScape: U.S. Crop Condition Monitoring Service, R.Mueller et al, AGU, Fall 2013.
[7] GEOGLAM Crop monitor: a geoglam initiative. www.geoglam-crop-monitor.org, Accessed 1 Mar 2017
[8] N. Murakami et al., Development of a teleoperation system for agricultural vehicles. *Comput. Electron. Agric.* 63(1):81-88, August 2008. ( http://dx.doi.org/10.1016/j.compag.2008.01.015 )

opportunity is to invest in research and development to leverage small satellites (e.g., Planet Labs[9]), which will provide high-resolution (e.g., daily global scans at 1meter resolution) to monitor crops for timely detection and management of adverse conditions. It is also important to invest in the modernization[10] of the Global Positioning System to protect against outage, jamming and spoofing as it is a crucial infrastructure for the precision agriculture during narrow time-windows for harvesting or planting large farms.

Broadly, the complex, interdisciplinary, and changing nature of agricultural processes and these urgent, potentially conflicting goals, demand technological breakthroughs that leverage recent advances in precision agriculture and scalable data management, while integrating new approaches for remote sensing and advanced analytics to provide effective, low cost, data-driven decision support and automation for the next-generation in farming operations. To enable such breakthroughs, an intelligent cyberinfrastructure must leverage recent technological advances that have proven successful for other sectors of the economy in spurring economic growth, profits, and high-paying job creation. One key example of this is e-commerce for the retail sector — in which companies such as Amazon, Netflix, and Walmart combine large-scale data analytics, complex modeling, and easy-to-use and scalable cloud systems to disrupt how consumers purchase goods and services. A similar disruption and impact is possible for the US agriculture industry through the use of these same underlying technologies tailored to food production processes in addition to the enhancement of citizen engagement and developing a private incentives regulatory framework to secure free access to data and quality of products delivered to users. Let us examine a few opportunities in detail in the following subsections.

*Table 1. Intelligent infrastructure and Research Needs*

| Areas | Intelligent Infrastructure and Research Needs |
|---|---|
| Workforce Development | Augmented reality: precision agriculture video-games to engage urban youth <br> Teleoperation: create jobs in labor-surplus areas & address farm labor shortage |
| Cyber Physical Systems & Robotics | Robust high-precision positioning to counter GPS[11] outage, jamming, & spoofing <br> Integrated sensors across satellites, UAVs, farming-equipment, and under-soil <br> Automation for labor intensive tasks, e.g., picking berries, pruning grape vines <br> Robotic bees for pollination in areas of declining bee population |
| Spatiotemporal Machine Learning, Data Analytics | Leverage new high-resolution (e.g., daily, 1 meter) satellite data to monitor crops <br> Spatiotemporal hotspot detection of agricultural pests, diseases and stresses <br> Model resource availabilities, forecast food, water & energy demands <br> Active management of sensors and actuators to optimize resource allocation |
| Security, Privacy, Safety | Secure, privacy-protected farm-data transmission and sharing spaces <br> Application-specific notions of privacy for data for spatiotemporal farm data <br> Economic models to promote data sharing among stakeholders |
| Networking, Internet of Farm Things | Improving Broadband Network Access in Rural Farming Areas <br> Edge Cloud Computing to reduce need for transferring large amounts of data |
| Decision Support | - Advanced spatiotemporal image, and video analysis techniques <br> Automate tasks e.g., identify crop stress, fruits/vegetables ready to be harvested |
| Citizen Engagement | Social Media, Apps, Easy to use Decision Support for growers and ranchers <br> Downstream behavioral change through apps (e.g., reduce food waste) <br> Cognitive and behavioral science applied to enhance feedback for technology improvement, scientific advancement and innovation |

## 4.1   Intelligent Workforce Infrastructure

Precision Agriculture Simulation and Tele-operation: To maintain and increase competitiveness of the U.S. agriculture industry in the global market, it is necessary to: (a) continuously retrain farmers and farm service providers so that they are well-positioned to embrace newer technologies; (b) engage next generation in the farming sector; (c) attract and train workers from other economic sectors (e.g., computer science, mechanical engineering, mining, manufacturing) for agriculture careers; and (d) go beyond workforce training to engage citizens and children. We need a multi-pronged approach. First, it is important to make agriculture "cool" in the minds of the next generation. This goal will need to be accomplished by highlighting modern technological tools such as drones, sensors, and robots and their applications in the agricultural sector as well as the gamification of agricultural processes. Nurturing virtual communities on popular social media platforms such as Facebook and Instagram will be vital in this regard. Second, small workshops and tutorials should be conducted to introduce current farmers as well as workers from other sectors to intelligent farming infrastructures and provide hands-on experience. Third, virtual and augmented reality platforms should be built to make the training more accessible, cost-effective and scalable. Associated video games may help engage young minds to aspire to be next generation farmers. Lastly, we should investigate tele-operation technologies to engage people in labor-surplus areas for jobs available on farms and farming areas. This approach may address the structural mismatch in the economy by employing people in labor surplus areas for unfilled and hard to fill jobs on farms.

---

[9] The Planet Labs, https://www.planet.com
[10] GPS.gov: GPS Modernization, http://www.gps.gov/systems/gps/modernization/
[11] GPS: The Global Positioning Systems, http://www.gps.gov

## 4.2    (Privacy-Protected) Shared Data Spaces

Although farmers are stewards of the land, they also must have sufficient revenue to stay in business; they must compete with their neighbors in commodity markets. As such, they may be reticent to share data about crops, soil, and equipment, since it provides a competitive advantage. Yet pooling data allows more powerful predictive analytics and optimization to accelerate adoption of best practices across farms to improve yield and farm profits while protecting water. Farmers are reluctant to share data, however, due to privacy concerns. There is a need to have new data analytic algorithms that maintain privacy. To maintain privacy when pooling data among neighboring farms to improve analytics, we have to recognize that data will be correlated. The soil in adjacent plots will be similar; the weather will also be common. The mathematical notion of *differential privacy* has emerged as a standard definition for preserving privacy through random perturbations when sharing information to optimize systems in a variety of industries ranging from health care (HIPAA) to education (FERPA). Unfortunately, enforcing such a universal definition significantly reduces utility of agricultural data due to spatiotemporal dependencies. Application-specific notions of privacy for data sharing are needed.

It is also necessary to build economic models that will promote sharing of data among multiple stakeholders, such as farmers, farm equipment manufacturers, co-operative societies and state and federal governments. Also note that agriculture is carried out to support human nutrition and so it is interlinked to the remainder of the food pipeline. Thinking in larger systems-level terms suggests possibly new computing approaches and optimizations not just in agriculture but in food manufacturing, retail, restaurants, and food waste management.

## 4.3    Intelligent Cyber-Infrastructure to support smart food systems

The intelligent agriculture cyberinfrastructure must integrate sensing (e.g., GPS, remote sensing, field sensors, etc.), data aggregation, scalable data analytics and visualization. Sensing will consist of stationary and mobile devices (e.g., smart phones, air/ground robots) that measure local environmental conditions (e.g., weather, soil moisture and composition), collect multi-spectral imagery (e.g., plant health, animal location, crop maturity), and track implement and input use (e.g., irrigation, pesticide, tractors) among others. Data systems will consist of public cloud services and on-farm or community-based edge cloud systems that implement a wide range of tools (e.g., open source and proprietary) for extracting actionable insights from farm data. Edge clouds are small computing "appliances" that operate similarly to public clouds yet preclude the need for Internet connectivity (and costly data transfer) while giving farmers real time, localized decision support and control over the privacy and sharing of their data. Public clouds will facilitate large-scale batch data analytics and sharing of anonymized information across farms. Networks link sensor and cloud systems to complete this end-to-end, multi-tier cyberinfrastructure that is key to enabling research and technology-transfer to the agriculture industry for open source and co-designed algorithms, programming environments, protocols, systems software, and analysis engines. Such systems are necessary to make it easy and economical to collect, mine, and analyze information (e.g., extracting inferences and predictions) and to form concrete solutions that can be directly tested, evaluated, and employed by U.S. growers to increase yields sustainably.

Finally, the data surrounding the crop life cycle and farming practices that such cyberinfrastructure must support is vast in size and disparate in type (e.g., imagery, time series, statistical), structure (e.g., hand-written, digitized), and scale (e.g., spatial and temporal, plant-to- global levels). Moreover, these data sets are incomplete, interdependent, volatile, imprecise, and generated by a vast diversity of devices (e.g., drones, farm workers, sensors, and Internet services) not designed to address future (and unknown) challenges. New techniques for data fusion are needed that integrate multi-dimensional data from multiple sources to form standardized and useful representation of a physical object or system that are amenable to analysis. Extracting actionable insights from this data requires new analyses that accurately describe, simulate, and model the complex systems that these data represent, such as the climate, soil, plant and animal genotypes and phenotypes, entomology, hydrological processes, human behavior, community food habits, and economic market forces, among others. Coupling models requires integrative techniques that preserve key system features and reduce uncertainty while eliding detail that results in additional computational workload without adding additional descriptive power. For agricultural productivity, these coupling techniques must be able to integrate analysis across scales (time and space) while reducing the number of dimensions of the problem at hand to enable computational tractability. Given the urgency of the problems that we face in this domain (e.g., the need for job creation, food safety, and significantly increased food production), such research and infrastructure must enable interdisciplinary collaboration and make available real-world test beds and data sets, from which validated results can be extracted and applied to the immediate problems facing both large and small holder farming concerns in the U.S.

**5.    Engaging the Computing Research Community in Smart Agriculture**

The computing research community may be engaged in this effort via community workshops to survey the current infrastructure for Food, Energy and Water as well as disruptive technology trends to identify research challenges and opportunities. The workshop may also help seed interdisciplinary partnership between computing researchers and Food, Energy and Water researchers. This may be followed by a research initiative with research funding and request for proposals for interdisciplinary research to envision and test feasibility of next generation infrastructure for Food, Energy and Water security.

**6. Acknowledgements**

We thankfully acknowledge the contributions of Jennifer Clarke, Roxanne Clements, Melissa Cragin, Travis Desell, Khari Douglas, Ann Drobnis, Wes Herche, Volkan Isler, Kyle Johnsen, Elizabeth Kramer, Vipin Kumar, Erin Mellinix, David Mulla, Luis Rodriguez, and Helen Wright.

*This material is based upon work supported by the National Science Foundation under Grant No. 1136993. Any opinions, findings, and conclusions or recommendations expressed in this material are those of the authors and do not necessarily reflect the views of the National Science Foundation.*